\documentclass{ws-procs9x6}

\usepackage{epsf,epsfig,subfigure,axodraw,graphicx,amsmath,amssymb}

\def\npb#1#2#3{Nucl.\ Phys.\ {\bf B#1} (#3) #2}
\def\plb#1#2#3{Phys.\ Lett.\ {\bf B#1} (#3) #2}
\def\prd#1#2#3{Phys.\ Rev.\ {\bf D#1} (#3) #2}
\def\prl#1#2#3{Phys.\ Rev.\ Lett.\ {\bf #1} (#3) #2}
\def\prp#1#2#3{Phys.\ Rep.\ {\bf #1} (#3) #2}

\def\jhep#1#2#3{JHEP\ {\bf #1} (#3) #2}

\begin{document}
\hfill{SNUTP 06-013}
\title{AXION: PAST, PRESENT AND FUTURE\footnote{Talk
presented at IDM 2006, Rhodes Island, Greece, Sep. 11--16, 2006.}}

\author{JIHN E. KIM}

\address{Department of Physics and Astronomy and\\
 Center for Theoretical Physics,\\
Seoul National University, Seoul 151-747, Korea \\
E-mail: jekim@phyp.snu.ac.kr}

\begin{abstract}
The current status of axion physics is presented. There still exists
the axion window $10^9$ GeV $\le F_a\le 10^{12}$ GeV. The recent
CAST solar axion search experiment on the axion-photon-photon
coupling strength has to be improved by a factor of 100 to reach
down to the region of superstring axions.  The calculable
$\bar\theta$ and $m_u=0$ cases for strong CP solutions, and axino
cosmology in SUSY extension of axion are also commented.
\end{abstract}


\bodymatter

%%%%%%%%%%%%%%%%%%%%%%%%%%%%%%%%%%%%%%%%%%%%%%%%%%%%%%%%%%%%%%%%%%%%%%
%%%%%%%%%%%%%%%%%%%%%%%%%%%%%%%%%%%%%%%%%%%%%%%%%%%%%%%%%%%%%%%%%%%%%%
\section{Introduction}

Modern cosmology needs dark matter and  dark energy in the universe:
$\Omega_{\rm CDM}\simeq 0.23, \Omega_\Lambda\simeq 0.73$. There are
several particle physics candidates for CDM: LSP,  axion, axino,
gravitino, LKP and other hypothetical heavy particles with some kind
of $Z_2$ symmetries.

The old electroweak scale axion is the pseudo-Goldstone boson
\cite{PQWW} arising from breaking the global Peccei-Quinn (PQ)
symmetry \cite{PQ}. The very light axion is the invention from the
need to solve the strong CP problem through PQ symmetry with
electroweak singlet field(s) \cite{veryaxion,kimprp87}. Superstring
axions \cite{Witten84,Witten85} may be in this very light axion
category. The existence of instanton solutions in nonabelian gauge
theories needs $\theta$ vacuum, introducing a CP violating
interaction \cite{CDGJR}. In the $\theta$ vacuum, the physically
meaningful interaction is parametrized by $\bar\theta$
$$
\frac{\bar\theta}{32\pi^2}F_{\mu\nu}\tilde F^{\mu\nu}\equiv\{F\tilde
F\};\quad \bar\theta=\theta_{\rm QCD}+\theta_{\rm weak}
$$
where $\theta_{\rm QCD}$ is the value determined from high energy
scale and $\theta_{\rm weak}={\rm Arg.\ Det.\ } M_q$ is the one
contributed when the electroweak CP violation is introduced. Here
$\bar\theta$ is the final value taking into account the electroweak
CP violation. For QCD to become a correct theory, this CP violation
by  $\bar\theta$ must be sufficiently suppressed. A nonvanishing
value ${\bar\theta}$ contributes to the neutron electric dipole
moment $d_n$. From the experimental limit \cite{Harris:1999jx}, we
obtain the bound
$$
 |d_n|< 0.63\times10^{25}\  e{\rm cm}  \to |{\bar\theta}|<
    10^{-9}.
$$
Why is this so small? It is the strong CP problem. There are three
types for the solution: (1) Calculable ${\bar\theta}$, (2) Massless
up quark, and  (3) Axion.

One may argue that there were no strong CP problem in the beginning.
In particular in 5D extension, since the instanton solution is the
one in 4D. I think this does not work or at best belongs to the
calculable $\bar\theta$ type, because in the 4D effective theory one
can always consider a 4D theory after integrating out the 5th
coordinate. Let us briefly comment on two solutions first.
\begin{itemize}
\item
The Nelson-Barr type \cite{NelBarr}: CP violation is introduced
spontaneously. So, original Yukawa couplings are real. Spontaneous
CP violation is introduced at high energy by introducing vectorlike
heavy quarks so that they mix with light quarks. If the {\it heavy}
vectorlike quarks are not introduced, the CP violation of light
quarks originated by the high energy scale CP violation will be tiny
due to the decoupling theorem. Not to be affected by the decoupling
theorem and to guarantee a tree level ${\rm Arg.\ Det.\ } M_q=0$,
specific forms for Yukawa couplings are assumed: SU(2)xU(1) breaking
real VEVs appear only between $F-F$ Yukawas, and
     CP violating phases in the VEVs appear only in $F-R$ Yukawas,
where   $F$ are the SM fermions and $R$ are the heavy fermions. If
heavy vectorlike fermions are integrated out, the effective Yukawa
coupling structure of the low energy sector is of the
Kobayashi-Maskawa form.
\item Massless up quark:
   Suppose that we chiral-transform a quark,
   $$
q\to e^{i\gamma_5\alpha}q.
   $$

It is equivalent to changing $\theta\to\theta-2\alpha$. Thus, if it
is allowed to have such a symmetry then strong CP problem is not
present. The massless quark case belongs here. This solution was
known from the very beginning of the strong CP problem but was not
taken seriously because the up quark seemed to be
massive\cite{Weinberg79}. The problem is, $\lq\lq$Is $m_u=0$ allowed
phenomenologically?"  The famous up/down quark mass ratio from
chiral perturbation theory (cPT) calculation is
$\textstyle\frac{m_u}{m_d}=\textstyle\frac59.$ But physics below 100
GeV is more involved. There is the determinental interaction of 't
Hooft, pictorially shown as
\begin{figure}[h]
\begin{center}
\begin{picture}(400,60)(0,0)
 \GCirc(160,30){10}{0.7}
 {\SetWidth{1}
 \LongArrow(100,30)(148,30)\Text(97,30)[r]{$d_L$}
 \LongArrow(105,60)(151,37)\Text(102,60)[r]{$u_L$}
\LongArrow(105,0)(151,23)\Text(102,0)[r]{$s_L$}
 \LongArrow(172,30)(220,30)\Text(223,30)[l]{$d_R$}
 \LongArrow(169,37)(215,60)\Text(218,60)[l]{$u_R$}
\LongArrow(169,23)(215,0)\Text(218,0)[l]{$s_R$}
 }
\end{picture}
\end{center}
\end{figure}
Below the electroweak scale quarks obtain mass. Suppose, the up
quark is massless. Then, there is no strong CP problem. But chiral
perturbation theory can be done with instanton generated  up quark
mass from the above 't Hooft interaction,
             $  m_u = m_d m_s / \Lambda$,
where $\Lambda$ is at the QCD scale. So it is the problem whether
the instanton calculus really gives the desired magnitude, in which
case $\theta$ is still unphysical. In the community, still there is
a disagreement on this issue:
     Kaplan and Manohar (KM), and Choi belongs to the positive
     group \cite{KM,Choi92}, and  Leutwyler (L) belongs to the negative
     group \cite{Leut}.
CP even observables do not see $m_u$. From the figure, for example,
we have $m_{u,eff}=m_d m_s / \Lambda,\ m_{d,eff}=m_d+m_u m_s /
\Lambda\simeq m_u m_s /\Lambda,\ m_{s,eff} \simeq m_s$. But CP odd
observables see $m_u$. Is $Z=m_u/m_d$ small? KM shows from the 2nd
order cPT  $Z\simeq 0.2$, and they could not rule out the $m_u=0$
case. Explicitly, cPT has the $L_7$ parameter in the term $L_7
\langle M^\dagger U-MU^\dagger\rangle$, where $M=3\times 3$ mass
matrix and $U=3\times 3$ matrix for meson fields. KM shows
\begin{align}
  m_u=0 :  {\rm cPT}\Rightarrow\left\{\begin{array}{l}
    L_7 \sim +1.5\times 10^{-4}\ {\rm or}\\
           (2L_8-L_5) \simeq (-1.2 \sim
          -2.5)\times 10^{-3}\end{array}\right.\label{KM}
\end{align}
where $L_8, L_5$ are another parameters in the cPT. On the other
hand L attempted to compute $L_7$, using the QCD sum rule for the
SU(3) singlet pseudoscalar $\eta^\prime$ dominance (similarly to the
vector meson dominance),
\begin{equation}
         L_7 \simeq L_{7,\eta^\prime} \simeq (-2 \tilde  -4)\times
         10^{-4}  (\rm Gasser-Leutwyler\
         coefficients)\label{Leut}
         \end{equation}
with a notable sign difference from (\ref{KM}).
 If (\ref{Leut}) were true, the case $m_u = 0$ is ruled out. But
Choi\cite{Choi92} argues that if $\eta^\prime$ gets mass from
instanton calculus, which is the modern wisdom on the U(1) problem
resolution, he can change the sign of (\ref{Leut}) to
\begin{equation}
  L_7 \simeq (3 \sim 8)\times 10^{-4}.
  \end{equation}
  So, we can have the possibility of $m_u = 0$.

In recent years,  lattice  calculation has been performed toward
this issue \cite{mulattice}. In $16^3 \times 32$ lattice
calculation, they obtain: $2L_8-L_5 \sim 10^{-4}$ and $ m_u /m_d =
0.484 \pm 0.027$.
        If true, $m_u=0$ is ruled out.

I consider that the problem on $m_u=0$ is not completely settled
yet, even though $m_u\ne 0$ seems to be the majority opinion of the
community.

\item
These show that the axion solution is the most compelling solution
which is discussed in the subsequent section.
\end{itemize}

%%%%%%%%%%%%%%%%%%%%%%%%%%%%%%%%%%%%%%%%%%%%%%%%%%%%%%%%%%%%%%%%%%%%%%%%%%%
%%%%%%%%%%%%%%%%%%%%%%%%%%%%%%%%%%%%%%%%%%%%%%%%%%%%%%%%%%%%%%%%%%%%%%%%%%%
\section{Axion}
    The axion potential is of the form
\begin{figure}[h]
\begin{picture}(400,15)(0,0)
{\SetWidth{1}\Curve{(55,10)(165,0)(275,10)}} \LongArrow(50,0)(280,0)
\Text(286,0)[l]{$a$} \Text(250,8)[c]{$\bullet$}
\end{picture}
\end{figure}
where the vacuum is shown as a bullet. The vacuum stays there for a
long time, and oscillates when the Hubble time ($1/H$) is larger
than the oscillation period($1/m_a$):
                       $   H < m_a $.
This occurs when the temperature is about 1 GeV. Axion is directly
related to $\theta$. Its birth was from the PQ symmetry whose
spontaneous breaking introduced a dynamical degree, a
pseudo-Goldstone boson called axion. But $\lq\lq$pseudo-Goldstone"
nature is specific in axion in that axion is a pseudoscalar $a$
without any potential except that arising from,
\begin{equation}
\frac{1}{32\pi^2} \frac{a}{F_a} F\tilde F\equiv\frac{a}{F_a} \{
F\tilde F\}.
\end{equation}
This kind of nonrenormalizable term can arise in several ways. The
first important scale is $ F_a$, defining the strength of
nonrenormalizable interaction. It can arise from higher dimensional
fundamental interactions with the Planck scale
$F_a$\cite{Witten84,Witten85}, from composite models with the
composite scale $F_a$ \cite{KimComp85}, from spontaneously broken
renormalizable field theories. In the last case, the global symmetry
must have the gluon anomaly and is called the PQ symmetry \cite{PQ}.
If this PQ symmetry is spontaneously broken, there arises a
pseudo-Goldstone boson \cite{PQWW} coupling to the anomaly with the
global symmetry breaking scale $F_a$.

In QFT, a very light axion is embedded in the phase of a complex
SU(2)$_L$ singlet scalar field $s$, (it may contain very tiny
components ($\le 10^{-7}$) from SU(2)$_L$ doublet phases)
\cite{veryaxion},
\begin{equation}
s=\frac{V+\rho}{\sqrt2}e^{ia/F_a},\quad a\equiv a+2\pi N_{DW} F_a
,\quad V= N_{DW}F_a.
\end{equation}
So, $F_a$ is in general smaller than $\langle s\rangle$. The
potential arising from the anomaly term after integrating out the
gluon field is the axion potential. Three properties of the axion
potential are known:
\begin{itemize}
\item[(i)] It is periodic with $2\pi F_a$ periodicity,
 \smallskip
\item[(ii)] The minima are at $\langle a\rangle=0,\ 2\pi F_a,\
4\pi F_a,\cdots $\cite{PQ,VW},
 \smallskip
\item[(iii)] A set of minima is identical, leaving to a few ($N_{DW}$)
distinct vacua \cite{DWnumber}.
 \end{itemize}
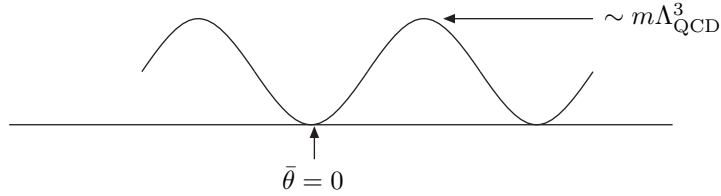
\begin{figure}[h]
\begin{center}
\begin{picture}(400,60)(0,0)
\Photon(70,40)(240,40){20}{2} \Line(20,20)(270,20)
\LongArrow(240,60)(185,60) \Text(245,60)[l]{$\sim m\Lambda^3_{\rm
QCD}$} \LongArrow(135,7)(135,18) \Text(135,0)[c]{$\bar\theta=0$}
\end{picture}
\end{center}
\caption{Vacua are at $\bar\theta=2n\pi$. The height of the axion
potential is given by the instanton interaction and boson mixing.}
\label{axpotential}
\end{figure}
The height of the axion potential is the scale $\Lambda$ of the
nonabelian gauge interaction and the boson mixing as shown in Fig.
\ref{axpotential}. We simply take this value as the QCD scale, but
in fact it is $m\Lambda_{\rm QCD}^3$ where $m$ is the light quark
mass \cite{Baluni}. The dominant one $\Lambda_{\rm QCD}^4$
corresponds to the $\eta'$ potential. If there are quarks, the
height is adjusted since as we have seen before a massless quark
makes it flat. The $u$ and $d$ quark phenomenology gives
\begin{equation}
V[a]= \frac{Z}{(1+Z)^2}f_\pi^2
m_\pi^2\left(1-\cos\frac{a}{F_a}\right).
\end{equation}
The essence of the axion solution is that $\langle a\rangle$ seeks
$\bar\theta=0$ in the evolving universe whatever happened before. It
is a cosmological solution \cite{axioncos} as shown in Fig.
\ref{axpotential}. The weak CP violation makes the minimum of the
potential shifted a little bit at $\bar\theta=O(10^{-17})$. The
axion mass is given by $m_a\simeq ({10^7\rm GeV}/{F_a})\ {0.6\ \rm
eV}$.

There are several laboratory experiments, restricting the axion
decay constants:
  (i) meson decays,
       $J/\Psi\to a+\gamma, \Upsilon\to a\gamma,
        K^+\to \pi^++a$,
  (ii) beam dump experiments,
       $p({\rm or\ }e-)N\to aX, a\to \gamma\gamma\
       {\rm and}\ e^+e^-$,
  (iii) and nuclear deexcitation,
       $N^*\to Na,  a\to  \gamma\gamma\ {\rm and}\ e^+e^-$.
Thus, we obtain the inner space bound $F_a\ge 10^4$ GeV from the
laboratory experiments. So, from the beginning, it was known that
the PQWW axion, arising from the electroweak scale, is ruled out
\cite{PecceiTok}. Thus, $F_a$ has to be very large, having led to
the so-called invisible axion. But, there is a possibility of
detecting it \cite{SikDet}, and hence it should be called a {\it
very light axion or sub-meV axion.}

%%%%%%%%%%%%%%%%%%%%%%%%%%%%%%%%%%%%%%%%%%%%%%%%%%%%%%%%%%%%%%%%%%%%%%%
%%%%%%%%%%%%%%%%%%%%%%%%%%%%%%%%%%%%%%%%%%%%%%%%%%%%%%%%%%%%%%%%%%%%%%%
\section{Axion window to outer space}

But the stringent lower bounds on the axion decay constant comes
from the outer space observations. Firstly, stellar evolutions, if
axion existed, are affected by axion emissions and the successful
standard energy loss mechanism due to weak interactions restricts
the axion mass toward a smaller region, or the axion decay constant
to a larger region. The stringent bound comes from the study of
supernova evolution \cite{supernova}, especially from the SN1987A
study to give $F_a\ge 10^9$ GeV \cite{SN1987Abound}. On the other
hand, the very interesting upper bound on $F_a$ is obtained from the
axionic contribution to dark energy in universe \cite{axioncos}.
%%%%%%%%%%%%%%%%%%%%%%%%%%%%%%%%%%%%%%%%%%%%%%%%%%%%%%%%%%%%%%%%%%%%%%%
\subsection{Stars}
The current supernova \cite{SN1987Abound} (globular cluster
\cite{glcluster}) limit on $F_a$ is $10^9$ GeV($10^{10}$ GeV). It
uses primarily the Primakoff process with the following coupling
\cite{kimprp87},
\begin{align}
{\cal L}_{a\gamma\gamma}=&-c_{a\gamma\gamma}\frac{a}{F_a}
             \frac{e^2}{32\pi^2}F_{\rm em}\tilde F_{\rm
             em}\Rightarrow{\bf E}\cdot{\bf B}\ {\rm interaction}\\
c_{a\gamma\gamma}=&~\tilde c_{a\gamma\gamma}+6\sum_{i=\rm light\
q}\tilde\alpha_iQ^2_{{\rm em},i}
\simeq\tilde c_{a\gamma\gamma}-1.93,\quad Z=\textstyle \frac59\\
 \tilde c_{a\gamma\gamma}=&~{\rm determined\ from\ high\ energy\
physics}\\
& \tilde\alpha_u\simeq\textstyle \frac{1}{1+Z},\quad
\tilde\alpha_d\simeq\frac{Z}{1+Z}
\end{align}
where the chiral symmetry breaking of $u,d$ quarks are taken into
account. The number 1.93 corresponds to $Z=\frac59$. Since the
instanton contribution to light quark masses is present
\cite{KM,Choi92}, we may take a band around 1.93.

   In the hot plasma in stars, once produced, they most probably escape
the core of the star and take out energy. This contributes to the
energy loss mechanism of star and should not dominate the
luminocity: (i) The Primakoff process:     $\gamma\to a$  (present
in any model): $g_{a\gamma\gamma}  < 0.6\times 10^{-10}\ {\rm
GeV}^{-1}$ or $F_a > 10^7$ GeV, and
                $0.4\ {\rm eV} < m_a < 200$ {keV ruled out
                because too heavy to produce,

(ii) Compton-like scattering: $\gamma e\to ae$ (DFSZ axion has $aee$
coupling)
                  $g_{aee} < 2.5\times 10^{-13},\
                       0.01\ {\rm eV} < m_a  < 200\ {\rm keV}$, and

(iii) SN1987A, $ NN\to NNa
     3\times 10^{-10} < g_{ aNN} < 3\times 10^{-7}  \Longrightarrow
       F_a > 0.6 \times 10^9\ {\rm GeV}$.

Stellar evolution uses the energy loss mechanism, with the
aforementioned lower bound on $F_a$. But laboratory experiments can
offer a more effective bound on $F_a$ than just the energy loss
mechanism, as done by the CAST(CERN axion solar telescope)
experiment of Fig. \ref{CASTexp} \cite{Zioutas:2004hi}.

\begin{figure}[t]
\centering \epsfig{figure=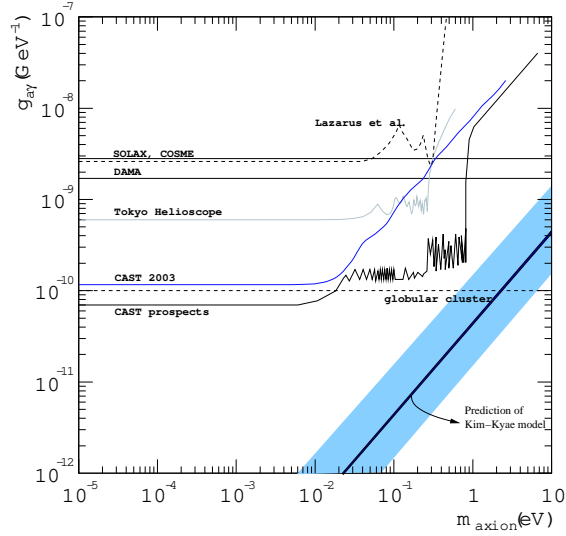, width=8cm}
\caption{\label{axionexp} Experimental bound from various
experiments, especially including CAST 2003 data. This figure is
taken from the result paper of CAST experiment
\cite{Zioutas:2004hi}. Here,we show the prediction from $Z_{12}$
model also. }\label{CASTexp}
\end{figure}

%%%%%%%%%%%%%%%%%%%%%%%%%%%%%%%%%%%%%%%%%%%%%%%%%%%%%%%%%%%%%%%%%%%%%%%
\subsection{Universe}

In the standard big bang cosmology (SBB), there is a severe domain
wall problem \cite{dwprob}. The SBB allows only the domain wall
number $N_{DW}=1$.\cite{BarrCK} But the most interesting
inflationary cosmology solves this domain wall problem at one stroke
if the reheating temperature after inflation is lower than $F_a$.
The inflationary cosmology seems to get support from the
            COBE and WMAP observations of density
            perturbations in the early universe,
            and NDW problem is not an issue since
in SUSY models the reheating temperature is required to be smaller
than $10^9$ GeV.\cite{reheat} In axion cosmology, the following
items are important:
\begin{itemize}
\item The axion decay constant $F_a$,
\smallskip
\item Axion couplings to $\gamma, e, p, n$,
\smallskip
\item The domain wall number $N_{DW}$.
\end{itemize}
If a singlet scalar VEV $\langle s\rangle$ breaks the PQ symmetry,
then $F_a=\langle s\rangle/N_{DW}$ defines the domain wall number.
Reheating temperature after inflation is required to be below $F_a$
if $N_{DW}>1$.

Axions are created at $T\simeq F_a$, but the universe does not
change $\langle a\rangle$ until $H\simeq m_a (T=1\rm GeV)$. Then,
the classical field $\langle a\rangle$ starts to oscillate. From the
harmonic oscillator type energy density $m_a^2 F_a^2$, we have $
 m_a \times {\rm number\ density} \Rightarrow$ CDM-like energy
  \cite{Turner86}:
\begin{equation}
\rho_a(T_\gamma)=m_a(T_\gamma) n_a(T_\gamma)\simeq
 2.5\frac{F_a}{M_P}\frac{F_am_a}{T_1}T_\gamma^3\left(
 \frac{A(T_1)}{F_a}\right)^2
\end{equation}
where the oscillation start-up temperature $T_1$ is the strong
interaction scale 1 GeV and $T_\gamma$ is the present temperature.
If $F_a$ is large($> 10^{12}$ GeV), then the axion energy density
dominates the energy density of the universe. Since the energy
density is proportional to the number density, it behaves like a
CDM.
%%%%%%%%%%%%%%%%%%%%%%%%%%%%%%%%%%%%%%%%%%%%%%%%%%%%%%%%%%%%%%%%%%%%%%%
\subsection{Axion window and search for cosmic axions}

The above astro- and cosmological-bounds on $F_a$ are summarized as
\begin{equation}
10^{9}{\rm GeV}\le F_a\le 10^{12}{\rm GeV}
\end{equation}

If axions are the CDM component of the universe, then they can be
detected. The feeble coupling can be compensated by a huge number of
axions. The number density $\sim F_a^2$, and the cross section $\sim
1/F_a^2$, and there is a hope of detecting it. Sikivie¡¯s cavity
detector \cite{SikDet} with dimension of tens of cm  has been used
to give coarse bounds \cite{cosaxsearch} on axion parameters in the
axion mass of order $10^{-5}$ eV.

%%%%%%%%%%%%%%%%%%%%%%%%%%%%%%%%%%%%%%%%%%%%%%%%%%%%%%%%%%%%%%%%%%%%%%%%%
%%%%%%%%%%%%%%%%%%%%%%%%%%%%%%%%%%%%%%%%%%%%%%%%%%%%%%%%%%%%%%%%%%%%%%%%%
\section{Axions from superstring}

Superstring tells us definite things about global symmetries. If
axion is present, it is better to be realized in superstring. They
are the bosonic degrees in $B_{MN}$ (MI-axion\cite{Witten84} is
$B_{\mu\nu}$ and MD-axion\cite{Witten85} is $B_{ij}$) and
furthermore additional massless bosons from compactification are
candidates. Superstring does not allow global symmetries. But there
is an important exception to this claim: the shift symmetry of
$H_{\mu\nu\rho}$, which gives the MI-axion. It is the only allowed
global symmetry. $B_{ij}$ are generally heavy \cite{WenWitten}; but
it is a model dependent statement.

The superstring axion decay constants are expected near the string
scale which is too large \cite{choikim85}:
            $ F_a  > 10^{16}$ GeV.
A key question in superstring models is $\lq\lq$How can one obtain a
low value of  $F_a$?" An idea is the following:
\begin{itemize}
\item[] In some compactifications, anomalous U(1)
results\cite{anomalU1}, where U(1) gauge boson eats the MI-axion to
become heavy. Earlier, this direction, even before discovering
anomalous U(1) gauge boson, was pointed out by Barr \cite{Barr85}.
It became a consistent theory after discovering the anomalous U(1).
Then, a global symmetry survives down the string scale. $F_a$ may be
put in the axion window.  It was stressed in several references
\cite{Kim88,Svrcek}.

\end{itemize}
However, this idea does not work necessarily, as will be commented
later.

Somehow MD-axion(s) may not develop a large superpotential terms.
But the problem here is the magnitude of the decay constant.
MD-axion decay constants were tried to be lowered by localizing them
at fixed points \cite{Conlon,ConWitKim}. It uses the flux
compactification idea and it is possible to have a small $F_a$
compared to the string scale as in the RS model. One needs the
so-called throat as schematically shown in Fig. \ref{s2s2s1}.
\begin{figure}[h]
\begin{center}
\epsfig{figure=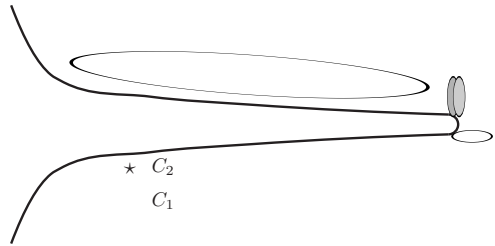, width=7cm, bbllx=275, bblly=650,
bburx=540, bbury=800}
\end{center}
\caption{A schematic view of the
$S_2\times S_2\times S_1$ throat. At the tip, one elongated $S_2$
can shrink to a point shown as a star. The un-shrunk $S_2$ has the
cycle $C_2$ where a MD-axion resides as a harmonic
2-form.}\label{s2s2s1}
\end{figure}

\paragraph{Axion mixing}

Even if we lowered some $F_a$, we must consider hidden sector also.
In this case, axion mixing must be considered. There is an important
theorem.

\begin{itemize}
\item[] {\it Cross theorem on potential heights and decay constants}
\cite{kim99}: Suppose two axions $a_1$ with $F_1$ and $a_2$ with
$F_2\ (F_1 \ll F_2)$ couples to two nonabelian groups whose scales
having a hierarchy, $ \Lambda_1\ll\Lambda_2$. The higher potential
$\Lambda_2$ couples to both axions. Then, the diagonalization
process chooses the larger potential $\Lambda_2$ corresponds to the
smaller decay constant $F_1$, and the smaller potential $\Lambda_1$
corresponds to the larger decay constant $F_2$.
\end{itemize}
So, just obtaining a small decay constant is not enough. Hidden
sector may steal the smaller decay constant. It is likely that the
QCD axion chooses the larger decay constant. Recently, the mixing
effect has been stressed by I.-W. Kim {\it et. al.}
\cite{ConWitKim,KSCIW}.

And most probably, our axion will couple to the $\mu$ term:
$$
                      H_uH_d f(S_1,S_2,\cdots).
$$

After all, the topologically attractive $B_{MN}$ may not be the
axion we want. Let us go back to earlier field theoretic very light
axion. In string models, its effect toward phenomenology was not
calculated before. Now we have an explicit model for MSSM
\cite{KimKyae2}, and we can see here whether the idea of approximate
global symmetry is realized. It must be that at sufficiently higher
orders the PQ symmetry is broken. In the $Z_{12-I}$ model, we
calculate the axion-photon-photon coupling \cite{KSCIW} whose result
is shown in Fig. \ref{CASTexp}. But the decay constant is at the GUT
scale. In this kind of calculation, there are so many Yukawa
couplings to consider. For example, we encountered O(104) terms for
d=7 superpotential terms and it is not a trivial task to find an
approximate PQ symmetry direction.

In addition, we point out that the MI-axion with anomalous U(1)
always has a large decay constant since most of the fields are
charged under this anomalous U(1). Phenomenologically successful
axion must need an approximate PQ symmetry.

An approximate PQ global symmetry with discrete symmetry in SUGRA
was pointed out long time ago: given by Lazarides and Shafi
\cite{LSDW} for a discrete  $Z_3\times Z_3$. But this field
theoretic method does not guarantee that string models realize this
idea. In this sense, an explicit demonstration of an approximate PQ
symmetry is vital for a string matter axion.

%%%%%%%%%%%%%%%%%%%%%%%%%%%%%%%%%%%%%%%%%%%%%%%%%%%%%%%%%%%%%%%%%%%%%%%%%
%%%%%%%%%%%%%%%%%%%%%%%%%%%%%%%%%%%%%%%%%%%%%%%%%%%%%%%%%%%%%%%%%%%%%%%%%
\section{SUSY extension and axino}

The SUSY extension always introduces gravitino. Gravitinos produced
thermally after inflation decay very late in cosmic time scale
($>10^3$ s) and can dissociate the light nuclei by its decay
products. Not to have too many gravitinos, the reheating temperature
must be bounded \cite{reheat,reheat1}, $ T_r\le 10^9{\rm GeV}\ ({\rm
old\ value}),\ T_r\le 10^7{\rm GeV}\ ({\rm new\ but\ model\
dependent\ value}) $. Therefore, in SUSY theories  we must consider
the relatively small reheating temperature, and the domain wall
problem does not matter in axion cosmology.

The SUSY extension with the strong CP solution via axion introduces
its superpartner {\it axino}. Its cosmological significance is in
that it can serve as keV range warm dark \cite{RTW} matter or GeV
range cold dark matter \cite{CKKR}. Let us comment on its CDM
possibility.

For axino to be CDM, it must be stable or practically stable.
Without the $R$-parity conservation, this can not happen. Thus, we
require the practical $R$-parity conservation for the possibility of
axino CDM. In addition, for axino to be LSP it must be lighter than
the lightest neutralino whose mass is expected to be around 100 GeV.
Thus, the estimation of the axino mass is of prime importance. The
conclusion is that there is no theoretical upper bound on the axino
mass and axino mass can be easily in the GeV range. Since axion is
almost massless, one expects that its superpartners are massless in
the first approximation. Its scalar partner, saxion, obtains the
mass of order the soft terms after SUSY breaking. Its cosmological
effect is relatively late decaying nature, adding more photons after
its decay \cite{KimSaxion}. Regarding mass, saxion  is like the
other SM SUSY scalars.

However, the axino mass is intimately related to the SUSY breaking
scenario and symmetries of the superpotential. The PQ symmetry
allows the following superpotential \cite{Kim83}
 $$
 W = f Z(S_1S_2-F_a^2),\quad    Z, S_1, S_2:\ \rm singlets
 $$
There also exist SUSY breaking soft terms.  Thus, the following
potential is obtained
$$
 V = |f|^2(|S_1|^2+|S_2|^2)|Z|^2+(A_1fS_1S_2Z-A_2fF_a^2Z+\rm h.c.)
$$
which determines the VEV of $Z$. Since $S_1$ and $S_2$ are of order
$F_a$, $\langle Z\rangle$ is of order the $A$ term. Thus, the
fermion partners have the mass matrix of the form
$$
\left(  \begin{array}{ccc} 0&m_{\tilde a} & fF_a\\
 m_{\tilde a}& 0& fF_a\\
  fF_a & fF_a& 0
 \end{array} \right),\quad
m_{\tilde a}=f\langle Z\rangle
$$
The lightest eigenvalue is the axino mass. The others are of order
$F_a$. As shown above, the axino mass is basically a free parameter,
and expected to be smaller than the naive SUSY breaking scale due to
the small coupling. Most probably, it is lighter than neutralino
$\chi$. But its mass can be much smaller than the SUSY scale as
shown from a superpotential of the form \cite{CKNaxino},
$$
                  W^\prime = f Z(S_1S_2-X^2) +\frac13(X-M)^3
$$
where $X$ carries the vanishing PQ charge. This potential is much
more complicated to analyze. We show
         $ m_{\tilde a} = O(A-2B+C) + O(m^2_{3/2}/F_a)$.
For the standard pattern of soft terms, we have $B=A-m_{3/2}$ and $
C=A-2m_{3/2}$ \cite{Nilles84}. Thus, the axino mass is of order keV.
Even the tree level axino mass needs the knowledge on the full
superpotential, we treat the axino as the LSP which is the most
probable choice. Its mass is left as  a free parameter. KeV axinos
can be warm dark matter which is thermal relic. GeV axinos can be
CDM. In this case, the reheating temperature must be low
\cite{CKKR}.

If gravitino is the next lightest LSP (NLSP), $ m_{\tilde
a}<m_{3/2}<m_{\chi} $, the gravitino problem can be resolved
\cite{Asaka}, since the thermally produced gravitinos would decay to
axino and axion which do not affect BBN produced light elements.

Most probably, $\chi$ would be the NLSP, and the thermal production
mechanism restricts the reheating temperature after inflation as
summarized in Fig. \ref{fig:TRaxino}.
\begin{figure}
%\centering
\epsfig{figure=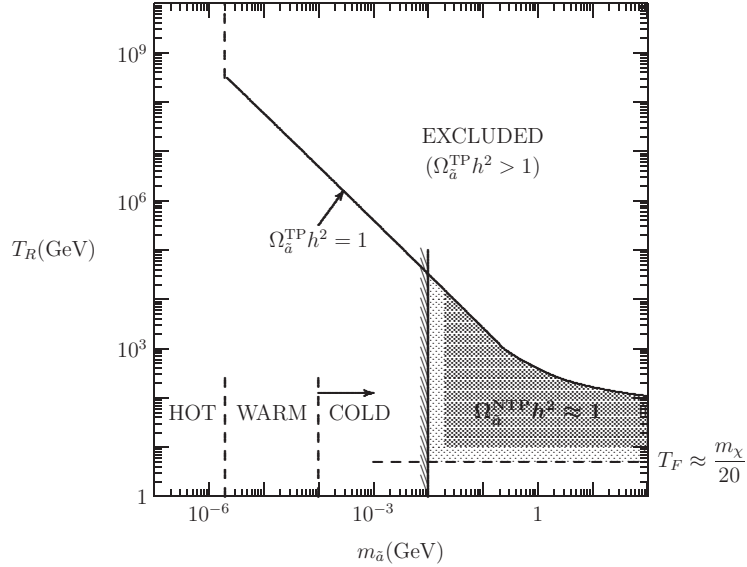, width=10cm,
  bbllx=90 ,bblly=380,  bburx =500 , bbury = 700}
\caption{The solid line gives the upper bound from thermal
production on the reheating temperature as a function of the axino
mass.  The dark region is the region where non-thermal production
can give cosmologically interesting results ($\Omega_{\tilde a}^{\rm
NTP} h^2\simeq1$).}\label{fig:TRaxino}
\end{figure}
At high reheating temperature, thermal production contributes
dominantly in the axino production. Even though the reheating
temperature is below critical energy density line, there still
exists the CDM possibility by the non-thermal production (NTP)
axinos. Covi {\it et. al.} shows \cite{CKKR}
$$
NTP:\quad\Omega_{\tilde a}h^2=\frac{m_{\tilde
a}}{m_{\chi}}\Omega_{\chi} h^2 \quad {\rm for}\ \ m_{\tilde
a}<m_{\chi}<m_{3/2}
$$
In Fig. , NTP axinos can be CDM for relatively low reheating
temperature $< 10$ TeV, in the region
$$
10\ {\rm MeV}< m_{\tilde a}< m_{\chi},\quad {\rm NTP\ axino\ as\
CDM\ possibility.}
$$

The shaded region corresponds to the MSSM models with
$\Omega_{\chi}h^2<10^4$, but a small axino mass renders the
possibility of axino closing the universe or just 30\% of the energy
density. If all SUSY mass parameters are below 1 TeV, then
$\Omega_{\chi}h^2<100$ and sufficient axino energy density requires
$$
m_{\tilde a}> 1\ {\rm GeV},\quad\left\{ \begin{array}{l} \rm Low\
reheating\ is\ good\ in\ view\ of \\
\rm the\ recent\ gravitino\ problem. \\
\rm But\ not\ good\ with\ leptogenesis. \end{array} \right.
$$

%%%%%%%%%%%%%%%%%%%%%%%%%%%%%%%%%%%%%%%%%%%%%%%%%%%%%%%%%%%%%%%%%%%%%%%%%
%%%%%%%%%%%%%%%%%%%%%%%%%%%%%%%%%%%%%%%%%%%%%%%%%%%%%%%%%%%%%%%%%%%%%%%%%
\section{Conclusion}

I reviewed strong CP and axion. In particular,

\begin{itemize}
\item Solutions of the strong CP problem: Nelson-Barr,  $m_u=0$, axion.
    Axion $a$ is the most attractive and plausible solution.
    \smallskip

\item Axions can contribute to CDM. Maybe solar axions are easier to
    detect. Most exciting is, it confirms instanton physics  by
    observation.
    \smallskip

\item  Tried to present a superstring matter axion coupling for the
first time. A QCD axion from superstring may be a window to string.
\smallskip

\item With SUSY extension, O(GeV) axino can be CDM.  It is
    difficult to detect this axino from the DM search, but possible
    to detect at LHC as missing energy.
\end{itemize}

%%%%%%%%%%%%%%%%%%%%%%%%%%%%%%%%%%%%%%%%%%%%%%%%%%%%%%%%%%%%%%%%%%%%%%%%%
%%%%%%%%%%%%%%%%%%%%%%%%%%%%%%%%%%%%%%%%%%%%%%%%%%%%%%%%%%%%%%%%%%%%%%%%%
\section{Acknowledgments}
This work is supported in part by the KRF ABRL Grant No.
R14-2003-012-01001-0. J.E.K. is also supported in part by the KRF
grants, No. R02-2004-000-10149-0 and  No. KRF-2005-084-C00001.

%%%%%%%%%%%%%%%%%%%%%%%%%%%%%%%%%%%%%%%%%%%%%%%%%%%%%%%%%%%%%%%%%%%%%%%%%
%%%%%%%%%%%%%%%%%%%%%%%%%%%%%%%%%%%%%%%%%%%%%%%%%%%%%%%%%%%%%%%%%%%%%%%%%

\end{document}